\documentclass[aps,prd,preprint]{revtex4-1}
\usepackage{graphicx}
\usepackage{braket}
\usepackage{amsmath}
\usepackage{amssymb}
\usepackage{dsfont}
\usepackage{hyperref}
\usepackage{interval}
\usepackage{bm}
\usepackage{picture}
\usepackage[x11names]{xcolor}

\def\draft{0}

\if\draft1
\usepackage[markup=underlined]{changes}
\else
\usepackage[final]{changes}
\fi

\intervalconfig {
	soft open fences ,
}

\newcommand{\su}{\uparrow}
\newcommand{\sd}{\downarrow}
\newcommand{\tr}{\mathrm{tr}}
\newcommand{\Ndof}{N_{d.o.f.}}
\newcommand{\invsqrt}[1]{\dfrac{1}{\sqrt{#1}}}

\begin{document}
\if\draft1 \listofchanges \fi
\title{Entanglement entropy in lattice theories with Abelian gauge groups}

\author{M. Hategan}
\email{hategan@ucdavis.edu}
\affiliation{University of California Davis, Davis CA 95616, U.S.A.}

\date{\today}

\begin{abstract}
We revisit the issue of the geometrical separability of the Hilbert space of 
physical states on lattice Abelian theories in the context of entanglement 
entropy. We discuss the conditions under which vectors in the Hilbert space, as 
well as the gauge invariant algebra, admit a tensor product decomposition with 
a geometrical interpretation. With the exception of pure gauge lattices with 
periodic boundary conditions which contain topological degrees of freedom, we
show that the Hilbert space is geometrically separable.
\end{abstract}

\maketitle
\section{\label{sec:intro0}Introduction}

Entanglement entropy in quantum field theories has received increased interest
in the past few decades. It has been shown that in many cases it satisfies 
an area law prompting questions of how it might be related to the 
Bekenstein-Hawking black hole entropy, which also satisfies an area law. 
Unfortunately, entanglement entropy is also UV divergent, thus requiring the
use of a cutoff. Lattice field theory is naturally equipped with such a cutoff,
making it a good fit for performing entanglement entropy calculations. Except 
that, while matter fields on the lattice are perfectly localized on lattice
vertices, gauge fields are represented by links connecting vertices, which,
by definition, have a spatial extent. To further complicate matters, 
physically measurable quantities must satisfy gauge invariance, and defining
a physical entanglement entropy requires the use of such gauge invariant
objects, whose structure is even more complex. In particular, it is believed
that degrees of freedom in the physical Hilbert space of gauge theories cannot
be divided into geometric bipartitions without sacrificing gauge invariance.
In this paper we show precisely how this can be done in the case of lattice
gauge theories with Abelian groups, while focusing on the particularly simple
group $Z_2$ for clarity. 

In Section~\ref{sec:intro}, we give a technical overview of some of the 
relevant literature and show the various issues surrounding the geometric 
separability of the Hilbert space of Abelian lattice gauge theories. 
Specifically, we show that two inseparability proofs lead to severe consequences 
that extend beyond gauge theories and can be equally applied to scenarios that
are otherwise thought to be geometrically separable. We also show that there is
no unique choice for what we call degrees of freedom and that the value of
the entanglement entropy can depend on that choice. 

Section~\ref{sec:dualities} shows how choices of degrees of freedom can be 
related using dualities and that imposing geometric symmetries on the degrees 
of freedom such that they can be interpreted as degrees of freedom in a field
theory can narrow down the number of choices. Symmetry arguments can then be 
applied to various scenarios to obtain minimally constrained gauge invariant
choices of degrees of freedom.

We proceed with a detailed analysis of the physical Hilbert space in $2+1$ 
dimensional $Z_2$ gauge theories with free boundary conditions in 
Section~\ref{sec:states-2d-gauge} and their algebra in 
Section~\ref{sec:algebra} and show specific examples for a minimal, 
two-plaquette lattice in Section~\ref{sec:two-plaq-lat}. We also show 
how gauge invariant density matrices and partial traces can be implemented on 
the physical Hilbert space, leading to a gauge-invariant entanglement entropy.
The work in Sections~\ref{sec:states-2d-gauge} and \ref{sec:algebra} is 
subsequently used as a basis for the analysis of other lattice configurations.
Section~\ref{sec:pbc} analyzes $2+1$ dimensional lattices with periodic 
boundary conditions, which exhibit both a global constraint and topological
degrees of freedom. We show that maintaining lattice symmetries requires the
inclusion of the global constraint in the calculations of entanglement entropy.
On the other hand, topological degrees of freedom cannot be factored in a pure
gauge theory. Lattices in $3+1$ dimensions are studied 
Section~\ref{sec:three-d}. They are characterized by the existence of local
constraints of a geometrical nature which, again, must also be taken into 
account if lattice symmetries are to be preserved. Lattice theories that couple 
gauge fields to bulk matter fields are discussed in Section~\ref{sec:matter}.
Coupling to matter fields simplifies the Hilbert space since states can be 
expressed as tensor products of independent electric states on the links and
independent matter states at the vertices. Edge charges are addressed in 
Section~\ref{sec:surface-charges} as a combination of pure gauge theory in the 
bulk and matter-coupled theory on the edges. 

We conclude with some remarks on extensions to other Abelian gauge groups and 
the limitations of this analysis.

\section{\label{sec:intro}Review of relevant literature}

A survey of the literature~\cite{buividovich_entanglement_2008,
Casini:2013rba,Radicevic:2014kqa,Aoki:2015bsa,Ghosh:2015iwa,Donnelly:2011hn} 
indicates that the commonly held belief is that the physical Hilbert space in 
pure Abelian gauge lattices is not geometrically separable. In an early paper 
on the topic by Buividovich et al.~\cite{buividovich_entanglement_2008} it is 
stated that the physical Hilbert space in pure $Z_2$ gauge lattices in $2+1$ 
dimensions does not admit a geometrical separation by assigning complementary 
sets of gauge links to regions. It is then concluded that the physical Hilbert 
space is not geometrically separable at all, and that defining an entanglement 
entropy requires embedding the Hilbert space in an extended space. The 
embedding procedure then results in a contribution to the entanglement entropy 
that is given by a Shannon term of the probability distribution of degrees of 
freedom on the boundary, which is proportional to the area of the boundary. 
This is taken as validation of the procedure, since it leads to an entanglement 
entropy that has an area law for any gauge theory, thus qualitatively matching 
the area law of black hole entropy~\cite{Srednicki:1993im}. 

We illustrate briefly why the inseparability
proof in~\cite{buividovich_entanglement_2008} is problematic. The proof goes as 
follows:
\newcommand{\Hs}{\mathcal{H}}
\newcommand{\Hst}{\tilde{\mathcal{H}}}
\newcommand{\Psit}{\tilde{\Psi}}
\newcommand{\Cs}{\mathcal{C}}
\newcommand{\Pc}{\mathcal{P}_\mathcal{C}}
\newcommand{\Hsp}[1]{\Hs_{\perp #1}}
\begin{quote}
Assume a Hilbert space $\Hst$ and a strict subspace of $\Hst$ generated by 
the projection operator $\Pc$ of a constraint $\Cs$, $\Hs_0 \subset \Hst, 
\Hs_0 = \Pc\Hst,$ and take a decomposition 
$\Hst = \Hst_A \otimes \Hst_B$ with $\Hst_{A, B}$ partially supporting the 
constraint $\Cs$. That is, there exist vectors $\Psit_A^c \in 
\Hst_A, \Psit_B^c \in \Hst_B$ such that $\Pc\Psit_{A, B}^c \ne \Psit_{A, B}^c$ 
and there exist orthogonal vectors $\Psit_A^0 \in \Hst_A, \Psit_B^0 \in \Hst_B, 
\Psit_{A, B}^0 \cdot \Psit_{A, B}^c = 0$ such that $\Pc\Psit_{A, B}^0 = 
\Psit_{A, B}^0$. Now, assume that there exists a decomposition 
$\Hs_0 = \Hs_A \otimes \Hs_B$ such that $\Hs_A \subseteq \Hst_A$ and 
$\Hs_B \subseteq \Hst_B$. It follows that any vector in $\Hs_A$ or $\Hs_B$ can 
be written as a vector in $\Hst_A$ or $\Hst_B$, respectively. Consider the 
states $\Psi^0 = \Psi_A^0 \otimes \Psi_B^0 = \Psit_A^0 \otimes \Psit_B^0$ and 
$\Psi^1 = \Psi_A^1 \otimes \Psi_B^1 = \Psit_A^c \otimes \Psit_B^1$ with 
$\Pc\Psit_A^0 = \Psit_A^0$, $\Pc\Psit_B^0 = \Psit_B^0$, $\Pc\Psit_A^c \ne 
\Psit_A^c $, $\Psit_B^0 \ne \Psit_B^1$. Then, the $\Cs$-invariant subspace 
$\Hs_0$ must also contain the vector $\Psi_2 = \Psi_A^1 \otimes \Psi_B^0 = 
\Psit_A^c \otimes \Psit_B^0$. However, $\Pc\Psi_2 = \Pc\Psit_A^c \otimes 
\Pc\Psit_B^0 = \Pc\Psit_A^c \otimes \Psit_B^0 \ne \Psi_2$. In other words, 
$\Psi_2$ does not satisfy the constraint, therefore we arrive at a 
contradiction.
\end{quote}

\begin{figure}[tbp]
\centering
\includegraphics[width=.30\textwidth]{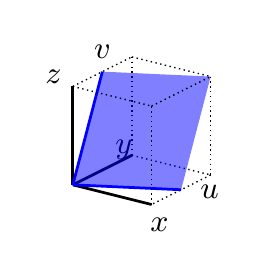}
\caption{\label{fig:cube-separability} Separability of non-trivial sub-spaces.
The full Hilbert space $\tilde{\Hs}$ is a 3-dimensional Cartesian 
space. The constrained space $\Hs_0$ (shaded) is a 2-dimensional 
space. It is impossible to express both $\hat{u}$ and $\hat{v}$ as vectors in 
any simple bipartition of $\{\hat{x}, \hat{ y}, \hat{z}\}$.}
\end{figure}

A simpler but somewhat inaccurate illustration of the problem is shown in 
FIG.~\ref{fig:cube-separability}.

The assumptions of the existence of the vectors satisfying the various 
relations above can be satisfied in lattice gauge theory. For example, 
$\Psi_{A,B}^0$ can be states with no electric excitations while $\Psi_A^c$ can 
be an open part of a closed electric string. There remains one assumption, 
$\Hs_0 = \Hs_A \otimes \Hs_B, \Hs_A \subseteq \Hst_A, \Hs_B \subseteq \Hst_B$,
which must be false. It must be noted, however, that this is not the same 
assumption as $\Hs_0 = \Hs_A \otimes \Hs_B$. Should this distinction not be 
made, one might be led to believe that no Hilbert space $\Hs$ is separable, 
since one can always find some $\Hs^* \supset \Hs$ and some constraint $\Cs^*$
with $\mathcal{P}_{\mathcal{C}^*} \Hs^* = \Hs$ such that the conditions in the 
proof are satisfied. There is, perhaps, some truth to this idea in that 
constraints can make the notions of degrees of freedom and locality ambiguous. 
One could write $\Hst = \Hs_A \otimes \Hs_B \otimes \Hsp{}$, where $\Hsp{}$ is 
the space of all vectors orthogonal to the constrained space $\Hsp{} = 
\ker \Pc$, and then attempt to factorize $\Hsp = \Hsp{A} \otimes \Hsp{B}$ such 
that $\Hst_{A, B} = \Hs_{A, B} \otimes \Hsp{A, B}$. This requires a meaningful 
assignment of the degrees of freedom in $\Hsp{A, B}$, which is not always 
possible: in the example in FIG.~\ref{fig:cube-separability}, $\Hsp{}$ is 
one-dimensional. This is likely the essence of the dilemma. Even when both the 
physical $\Hs_0$ and unphysical $\Hst$ spaces can be factored, a decomposition 
of the form $\Hst = (\Hs_A \otimes \Hsp{A}) \otimes (\Hs_B \otimes \Hsp{B})$ 
with $\dim(\Hsp{A}) = \dim(\Hsp{B})$ or $\dim(\Hsp{A}) / \dim(\Hs_A) = 
\dim(\Hsp{B}) / \dim(\Hs_B)$ may not exist.

\newcommand{\Al}{\mathcal{A}}
\newcommand{\Nv}{{\bar{V}}}
\newcommand{\Id}{\mathds{1}}
Casini et al.~\cite{Casini:2013rba} expand on the work 
in~\cite{buividovich_entanglement_2008} by looking at the problem from an
algebraic perspective. They conclude that expressing a constrained Hilbert 
space as a product space depends on the method in which one associates a region 
of space with an algebra of operators. In principle, a factorization of a 
Hilbert space $\Hs = \Hs_A \otimes \Hs_B$ is associated with a factorisable
algebra $\Al = \Al_A \otimes \Al_B$ such that $O_{A,B} \Psi_{B, A} = 
\Psi_{B, A},\, O_{A, B} \in \Al_{A, B},\, \Psi_{A, B} \in \Hs_{A, B}$. That is, 
operators $O_{A,B}$ have the form $O_A = \tilde{O}_A \otimes \Id_B$ and 
$O_B = \Id_A \otimes \tilde{O}_B$, where $\Id_{A, B}$ act trivially on their 
respective subspaces. A necessary and sufficient condition for the above 
factorization to exist is $\Al_A \cap (\Al_A)' = \Id$, where $(\Al_A)'$ is the 
set of all operators in $\Al$ that commute with \emph{all} operators in $\Al_A$.
The intersection $\Al_A \cap (\Al_A)'$ is the ``center'' of the algebra 
$\Al_A$. It is then claimed that lattice gauge theories do not admit local 
algebras with trivial center and therefore no factorization. The reasoning is:

\begin{quote}
	Consider, again, the earlier Hilbert space $\Hst$ and a constrained 
	subspace $\Pc\Hst = \Hs_0$. There exist operators $T^c$ that act 
	trivially on $\Hs_0$ but not on $\Hst$. That is, $\exists \Psit \in \Hst$ 
	such that $T^c\Psit \ne \Psit$ and $\forall \Psi \in \Hs_0, T^c\Psi = 
	\Psi$. There is an algebra $\Al$ associated with $\Hs_0$. Assume that there
	exists a factorization of $\Al = \Al_A \otimes \Al_B$ and that there exists
	a $T^c = A B$ with $A \in \Al_A, B \in \Al_B,$ and $B \ne \Id_B$. The 
	operator $B$ commutes with all operators in $\Al_A$ by the factorization 
	assumption. Then, since $T^c = \Id_{AB}$ on $\Hs_0$, we can write 
	$A^\dagger = A^\dagger\Id_{AB} = A^\dagger T^c = A^\dagger AB = B$. Given 
	$A^\dagger \in \Al_A$ then also $B \in \Al_A$. Hence $B \in \Al_A \cap 
	(\Al_A)' \ne \Id$ and the sub-algebra $\Al_A$ is not a factor, which 
	contradicts the factorization assumption.
\end{quote}

Perhaps the bigger problem would be that we are led to the conclusion 
that, in a bosonic theory, either the canonical conjugate to $A^\dagger$ is not
in $\Al_A$ or that it commutes with $A^\dagger$. That is unless 
$A^\dagger \Psi_A = \Psi_A, \forall \Psi_A \in \Hs_A$, in which case 
$A^\dagger$ is a trivial operator on $\Hs_A$ and it has no conjugate. The 
contradictions disappear if, from $T^c = \Id = AB$, one concludes that 
$A^\dagger \equiv B$ and that 
$T^c = (\Id_A \otimes B^\dagger)(\Id_A \otimes B)$ on the constrained subspace 
$\Hs_0$.

Imposing the requirement of a separable algebra, while sufficient, may not be 
necessary. Consider a two-spin system with the constraint 
$\sigma_z^1\sigma_z^2 = 1$. In other words, the Hilbert space is restricted 
to wave functions that satisfy 
$\sigma_z^1\sigma_z^2\ket{\Psi} = \ket{\Psi}$. By applying the constraint 
equation to a general state, we find that 
$\ket{\Psi} = \alpha\ket{\su\su} + \beta\ket{\sd\sd}$. This implies that we are 
not free to individually manipulate the spins and the algebra associated with 
this Hilbert space is $\Al = \{\mathds{1}, \sigma_z^1 = \sigma_z^2, 
\sigma_x^1\sigma_x^2,\, [\sigma_z^1, \sigma_x^1\sigma_x^2]\}$. Nonetheless, 
this can represent a legitimate Bell-type experiment, where two spatially 
separated observers can measure spin correlations and the following 
entanglement entropy:
\begin{align}
	\label{eq:sa-single-spin}
	S_A = -\alpha^2 \log \alpha^2 - \beta^2 \log \beta^2. 
\end{align}

This setup can be seen as either an entangled state or a single spin, depending 
on whether $\sigma_z^1$ and $\sigma_z^2$ are interpreted as distinct operators 
that measure the same quantity or distinct labels on the same operator. The 
distinction, however, is not algebraic. Instead, it hinges on whether 
independent physical measurements can be performed using measurement devices 
that have a clear spatial separation. In a field theory where the separation is 
near the scale cutoff, the notion of a clear separation disappears. Furthermore, 
this is precisely a Hilbert space that exhibits a gauge-like symmetry. To see 
this, we switch to the transverse basis using $\ket{\pm} = (\ket{\su} \pm 
\ket{\sd}) / \sqrt{2}$ and re-write the state in the new basis:
\begin{align}
	\label{eq:simple-state-el}
	\ket{\Psi} = 
		\dfrac{\alpha + \beta}{2}\left(\ket{++} + \ket{--}\right) -
		\dfrac{\alpha - \beta}{2}\left(\ket{+-} + \ket{-+}\right),
\end{align}
which is a state that is invariant under global transformations
$\ket{\pm} \rightarrow \ket{\mp}$. 

It can be noted that the algebra $\Al$ is
already invariant under this particular symmetry, so no additional information 
on the algebra can come from imposing invariance under this symmetry. We could
simply stop here noting that the value of $S_A$ in Eq.~(\ref{eq:sa-single-spin}) 
is precisely the classical Shannon entropy term found 
in~\cite{buividovich_entanglement_2008,Donnelly:2011hn} and that this choice
corresponds to a separable constrained space $\Hst$. However, as Casini et 
al. note in~\cite{Casini:2013rba}, $S_A$ would not be the same when gauge 
fixing is involved. We can remove the redundancy in 
Eq.~(\ref{eq:simple-state-el}) by selecting a particular point in the orbit to
get:
\begin{align}
	\ket{\Psi} = \dfrac{\alpha + \beta}{2}\ket{++} -
		\dfrac{\alpha - \beta}{2}\ket{+-} = 
		\ket{+}\otimes\left(\dfrac{\alpha + \beta}{2}\ket{+} -
		\dfrac{\alpha - \beta}{2}\ket{-}\right),
\end{align}
which is a separable state with $S_A = 0$ and corresponds 
to an entanglement entropy calculated on the unconstrained space $\Hs_0$. This 
is consistent with the idea that our Hilbert space contains a single degree of 
freedom.

We could, therefore, adopt the view that a degree of freedom is the smallest
entity that can be both measured and manipulated 
independently~\cite{Zanardi:2004zz} while respecting required symmetries. 
Unfortunately, this too can fail to result in an unambiguous entanglement 
entropy when multiple choices of basis have equally good geometrical 
interpretations. Consider the case of four spin degrees of freedom, one global 
constraint of the form $\prod_i \sigma_z^i = 1$ and the state
\begin{align}
	\ket{\Psi} = \invsqrt{2}\left(\ket{\su_1\su_2\su_3\su_4} + 
		\ket{\sd_1\su_2\sd_3\su_4}\right).
\end{align}

The space can be divided by assigning the first two spins to a region and the
other two to its complement. If we consider spin $1$ as redundant, the 
unconstrained state reads:

\begin{align}
	\ket{\Psi'} = \invsqrt{2}\ket{\su_2} \otimes \left(\ket{\su_3\su_4} + 
		\ket{\sd_3\su_4}\right).
\end{align}

Being a separable state, the entanglement entropy is zero. However, choosing
spin $4$ as redundant, we get:

\begin{align}
	\ket{\Psi''} = \invsqrt{2}\left(\ket{\su_1\su_2} \otimes \ket{\su_3} + 
		\ket{\sd_1\su_2} \otimes \ket{\sd_3}\right),
\end{align}
and the entanglement entropy is now $\log 2$. In one dimension, there is a 
simple solution to the problem which involves a change of basis to 
eigenstates of products of neighboring $\sigma_z$ operators. This solution 
preserves homogeneity of the degrees of freedom. In two dimensions such a 
solution does not exist. The resulting ambiguity of $S_A$ is endemic to spaces 
with global constraints in all but a few cases and stems from the lack of a 
unique way of meaningfully assigning coordinates to the unconstrained degrees 
of freedom that preserve various qualities that one would expect from a field 
theory. As we will show in Section~\ref{sec:states-2d-gauge} certain theories 
admit duals with unconstrained degrees of freedom that can be interpreted as 
local field theories while others may not (Section~\ref{sec:three-d}).

We end this introduction by mentioning a concern introduced by Donnelly
in~\cite{Donnelly:2011hn}: edge states in gauge 
theories~\cite{Balachandran:1994vi}. This is based on earlier work by 
Witten~\cite{Witten1979} and Lowenstein and Swieca~\cite{LOWENSTEIN1971172}.
Lattice theories with edge states are special in that they are part of a class 
of theories that do not fully preserve the homogeneity of degrees of freedom 
from the outset.

\section{\label{sec:dualities}Degrees of Freedom and Dualities}

\newcommand{\sigmap}{\tilde{\sigma}_\square}

We have seen that entanglement can depend on the precise definition of what
a degree of freedom is and that there is generally no unique choice of degrees 
of freedom. The ambiguity is not necessarily specific to gauge theories, but to 
spaces with constraints. We will attempt to address two questions. One is 
whether there exist gauge invariant degrees of freedom that can be deemed as 
defining a discretized field theory and the second is whether there exists
a set of assumptions that can lead to an unambiguous entanglement entropy when
constraints are present.

In general, choices of degrees of freedom are related by dualities. Given a 
field theory defined on a discrete lattice with discrete Abelian group
valued degrees of freedom there exist dualities that preserve the type and 
number of unconstrained degrees of freedom. A way of constructing such 
dualities consists of finding a basis in the Hilbert space of the theory, a set 
of distinct operators $\hat{O}_i$, one for each degree of freedom, that are 
simultaneously diagonalized by the basis states, and then selecting some set of 
operators from the group generated by $\hat{O}_i$ under multiplication to 
define the new degrees of freedom. For a theory with constraints, we ask the 
question of whether we can find suitable dualities that have no constrained 
degrees of freedom.

Consider a simple example: a 2-d quantum Ising lattice with degrees of freedom 
on the vertices and a global constraint. The Hilbert space has 
$\Ndof = \left[(L_x + a)(L_y + a)/a^2 - 1\right]$ $Z_2$ degrees of freedom, 
where $a$ is the lattice spatial dimension. The $+a$ part can be seen as a 
matter of convention. One can enlarge the lattice by $a/2$ on all sides 
($L_i' = L_i + a$) and consider the degrees of freedom to be associated with 
the centers of plaquettes of the enlarged lattice and then write 
$\Ndof = N_x' N_y' - 1$, with $N_i' = L_i' / a$. We can also write 
$\Ndof = (N_x + 1)(N_y + 1) - 1 = N_x N_y + (N_x + N_y)$. This suggest that our 
theory could have a dual with $N_x N_y$ bulk degrees of freedom and $N_x + N_y$ 
edge degrees of freedom. To see that this duality exists, we follow the 
geometry suggested by the degree of freedom decomposition and associate with 
the center of every plaquette a degree of freedom defined by the following 
operator identities:
\begin{align}
	\sigmap^z(x + 1/2, y + 1/2) = 
	  \sigma^z(x, y)\sigma^z(x + 1, y)\sigma^z(x, y + 1)\sigma^z(x + 1, y + 1),
\end{align}
where we switched to lattice units $a = 1$. Similarly for edge degrees of 
freedom:
\newcommand{\sigmat}{\tilde{\sigma}}%
\begin{align}
	\sigmat^z_{E_x}(x + 1/2) &= \sigma^z(x, 0)\sigma^z(x + 1, 0) \nonumber\\
	\sigmat^z_{E_y}(y + 1/2) &= \sigma^z(0, y)\sigma^z(0, y + 1).
\end{align}

The algebra generated by all $\sigmap^z$ and $\sigmat^z_{E_d}$ will contain all
products $\sigma^z(x_1, y_1)\sigma^z(x_2, y_2)$, which are all simultaneously
invariant under a global spin flip. The dual theory does not contain the global
constraint and it may seem that we should prefer the dual in calculations 
where geometric ambiguities in the choice of basis are relevant. The problem 
with the dual, however, is that the standard nearest-neighbor terms in the 
Hamiltonian are non-local. To see this, consider the term 
$\sigma^z(x_0, y_0)\sigma^z(x_0, y_0 + 1)$. In terms of dual degrees of 
freedom, it takes the form:
\begin{align}
	\sigma^z(x_0, y_0)\sigma^z(x_0, y_0 + 1) = \sigmat^z_{E_y}(y_0 + 1/2) 
	\prod_{x = 0}^{x_0 - 1} \sigmap^z(x + 1/2, y_0 + 1/2).
\end{align}
\begin{figure}[tbp]
\centering
\includegraphics[width=.35\textwidth]{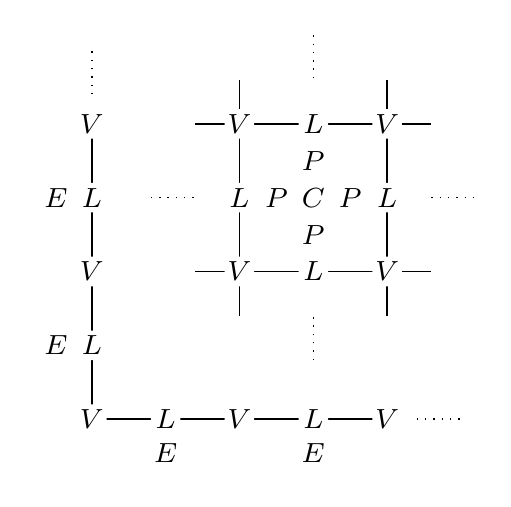}
\caption{\label{fig:unit-cell} Possible arrangements of degrees of freedom
on a square lattice.}
\end{figure}

Imposing a locality condition on the action would exclude the above duality 
from consideration. We note, however, that the dual has the exact same physics
content and the same algebra as the initial theory. The choice of one or the 
other is a matter of preference and this preference must be informed by other
considerations. It may then be desirable to consider other geometric symmetries 
that should be satisfied by a duality, such as homogeneity of the degrees of 
freedom, (discrete) isotropy, parity transformations, boundary conditions, as 
well as the preservation of the lattice spacing unit. For example, in two 
dimensions, only square lattices satisfy the 4-fold isotropy condition. This 
restricts the number of unconstrained degrees of freedom that can be 
represented on a square lattice with open boundary conditions 
(see FIG.~\ref{fig:unit-cell}):
\begin{align}
	\label{eq:ndof1}
	\Ndof^d &= (C + 4 P) N_{\text{plaq.}}^d + V N_{\text{vert.}}^d + 
		L N_{\text{links}}^d + E N_{\text{edge links}}^d \nonumber\\
		  &= (C + 4 P + 2 L + V) N_x^d N_y^d + (L + 2 E + V) (N_x^d + N_y^d) + 
		  V,
\end{align}
with $C, P, V, L, E \in \mathbb{N}$ being the number of degrees of freedom
associated with, respectively, the center of plaquettes, the off-center of
plaquettes, vertices, links, and edge links (and/or edge vertices). One can 
then check numerically that for a 2-d Ising model with a global symmetry there 
is no satisfactory duality by finding the intersection of sets of solutions
$\{(C, P, V, L, E)\}$ to the equation $\Ndof^d = \Ndof = N_x^0 N_y^0 - 1$ 
for various $N_x^0$ and $N_y^0$ which are sizes in the original model. In 
particular, one can verify that $\Ndof = 2\times 2 - 1 = 3$ has no solution 
except the trivial $C=3$ solution which places all degrees of freedom at the
center of one plaquette. The existence of solutions to the above equation does 
not necessarily imply that a suitable duality exists, but the absence of 
solutions implies the non-existence of an unconstrained duality satisfying the 
geometric symmetries.

It follows that eliminating constraints can, in certain cases, lead to theories 
that do not have a suitable interpretation as local field theories. In such 
cases it would seem that we must either accept the constraints as physical or 
give up locality or other symmetries.

\section{\label{sec:states-2d-gauge}Gauge invariant states in pure $Z_2$ gauge 
lattices}

We proceed with an analysis of the physical Hilbert space in $Z_2$ pure gauge 
lattices in $2+1$ spatial dimensions with free boundary conditions and a 
temporal gauge, and consider a time slice with all links in the spatial 
dimensions. This is the basic setup used 
in~\cite{buividovich_entanglement_2008} and it is a natural choice for 
a Hamiltonian lattice 
theory~\cite{creutzHamiltonianLattice,kogutSusskindLattice}. When working with 
a Wilsonian theory, one must also consider plaquettes with a time component 
which correspond to the electric components of the electromagnetic tensor. The
two are related, in the continuum limit, by 
$E_i^2 = F_{i0}^2 \approx (1/a^4)(1 - \mathrm{Re}\, U_{i0})$. The inclusion of 
electric plaquettes in the Wilsonian theory will be discussed as a particular 
case of a three-dimensional time slice in Section~\ref{sec:three-d}. For the 
Hamiltonian version of the theory, throughout the paper, we will assume the 
following Hamiltonian:
\begin{align}
	\label{eq:hamiltonian}
	H = \sum_{x, \mu} L_\mu(x) - \lambda \sum_{x, \mu, \nu} U_\mu(x) U_\nu(x)
		U_\mu(x + \hat{\nu}) U_\nu(x + \hat{\mu}), 
\end{align}
where the first sum is taken over all the links and the second is taken over
all the plaquettes. This corresponds to the following Wilson action:
\begin{align}
	\label{eq:action}
	S = -\lambda \sum_{x, \mu, \nu} U_\mu(x) U_\nu(x) U_\mu(x + \hat{\nu}) 
		U_\nu(x + \hat{\mu}).
\end{align}

Consider a lattice with $Z_2$ links and the standard $Z_2$ algebra of operators 
acting on the links:
\begin{align}
	\label{eq:ul-def}
	U \ket{u} &= u\ket{u} \\
	L \ket{u} &= \ket{-u},
\end{align}
with $u \in \{+1, -1\}$ (or, for consistency with spin systems, 
$u \in \{\su, \sd\}$), and the commutation relations:
\begin{align}
	[U_\mu(x), U_\nu(y)] &= 0 \\
	[L_\mu(x), L_\nu(y)] &= 0 \\
	[U_\mu(x), L_\nu(y)] &= 0,\, x \ne y \lor \mu \ne \nu\\
	[U_\mu(x), L_\mu(x)] &\ne 0.
	\label{eq:link-op-comm}
\end{align}

Gauge transformations are operators parametrized by group elements associated
with each vertex which transform links as follows:

\begin{equation}
	u_\mu(x) \rightarrow g(x) u_\mu(x) g^{\dagger}(x + \hat{\mu}),
\end{equation}
where $g(x)$ and $g(x + \hat{\mu})$ represent the vertices associated with the 
endpoints of link $u_\mu(x)$. In the $Z_2$ case, $u_\mu(x)$ is flipped if 
exactly one of $g(x)$ and $g(x + \hat{\mu})$ are $-1$. 

An arbitrary gauge transformation acts on an arbitrary link polynomial as 
follows:
\begin{align}
	u_{\mu_1}(x_1) u_{\mu_2}(x_2) ... u_{\mu_n}(x_n) \rightarrow 
		g(x_1) u_{\mu_1}(x_1) g^{\dagger}(x_1 + \hat{\mu_1})
		g(x_2) u_{\mu_2}(x_2) g^{\dagger}(x_2 + \hat{\mu_2}) \ldots \nonumber\\
		g(x_n) u_{\mu_n}(x_n) g^{\dagger}(x_n + \hat{\mu_n}).
\end{align}

This polynomial is gauge invariant only if all of the gauge terms cancel out, 
which can only happen for closed paths (Wilson loops), products of closed 
paths (for Abelian groups), or a constant. The smallest Wilson loop is the one 
that goes around a single plaquette.

This implies that we can use Wilson loop functionals to construct functionals 
$\Psi[U_\mu(x)] = \sum c_k W_k[U_\mu(x)]$ that result in gauge invariant states:
\begin{align}
	\ket{\Psi} 
		&= \sum_{u_\mu(x) = \pm 1} \Psi[U_\mu(x)] 
			\bigotimes_{\mu, x} \ket{u_\mu(x)} 
		 = \sum_{u_\mu(x) = \pm 1} 
			\sum_{k} c_k W_k[U_\mu(x)] \bigotimes_{\mu, x} \ket{u_\mu(x)},
\end{align}
where $W_k[U_\mu(x)]$ are any subset of the Wilson loop polynomials, including 
the identity. These states are gauge invariant because they assign the same 
coefficients to all microstates $\bigotimes_{\mu, x} \ket{u_\mu(x)} $ related 
by gauge transformations.

\begin{figure}[tbp]
\centering
\includegraphics[width=.70\textwidth]{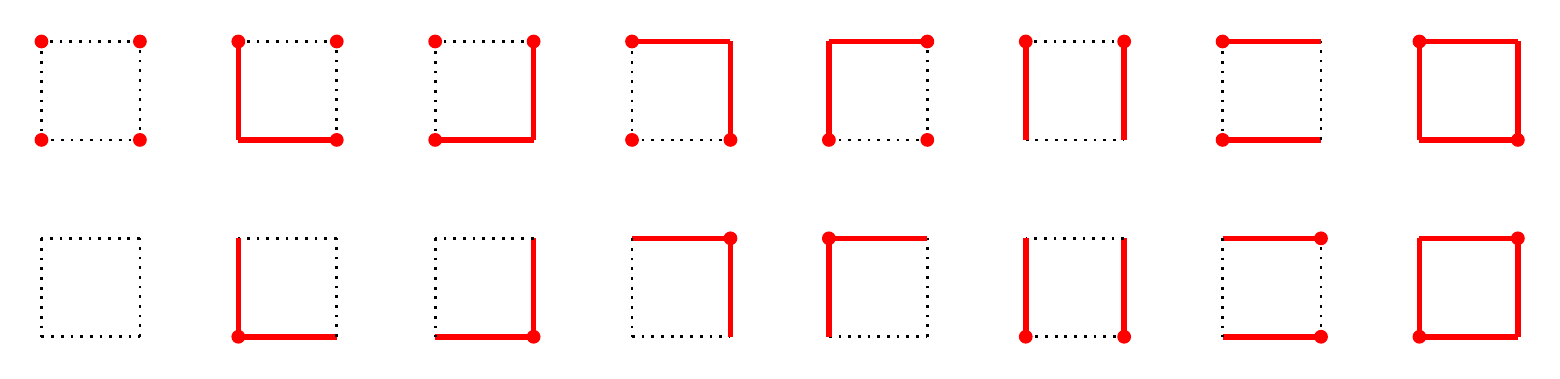}
\caption{\label{fig:plaquette-gt} Gauge transformations on a $Z_2$ plaquette. 
The gauge transformations happen at the marked vertices. The solid links are 
links that are affected by the gauge transformation. Transformations in the 
same column are equivalent.}
\end{figure}

For a single plaquette, the gauge transformations can be seen explicitly 
in FIG.~\ref{fig:plaquette-gt}. Since there are $2^4 = 16$ total link states
and $8$ distinct gauge transformations, there are exactly $2$ physical states
per plaquette. A gauge transformation on a $Z_2$ lattice will always flip an 
even number (including none) of links in any given plaquette. Consequently, a 
convenient basis for physical states is obtained by dividing the kinematic 
(link) states into states with even/odd number of up-type links per plaquette. 
These states are eigenstates of plaquette operators:
\begin{align}
	U_{\mu\nu}(x) = U_\mu(x) U_\nu(x + \hat{\mu}) 
		U^\dagger_\mu(x + \hat{\nu}) U^\dagger_\nu(x).
\end{align}

\newcommand{\UP}{U_{\square}}
In two dimensions, we can drop the tensor indices and use 
$\UP(x) \equiv U_{12}(x)$. The states can be constructed using products of the 
operators
\begin{align}
	\UP^\su(x) &\equiv \invsqrt{2} (1 + \UP(x)) \\
	\UP^\sd(x) &\equiv \invsqrt{2} (1 - \UP(x)),
\end{align}
by applying them to the weak coupling ground state:
\begin{align}
	\ket{0} = C\sum_{u_\mu(x) = \pm 1} \bigotimes_{\mu, x} \ket{u_\mu(x)},
\end{align}
where $C$ is a normalization constant. Specifically, for a single plaquette, 
we have that
\newcommand{\sqsu}{\su_\square}
\newcommand{\sqsd}{\sd_\square}
\begin{align}
	\ket{\sqsu} &= \UP^\su\ket{0} = 
		\dfrac{1}{2\sqrt{2}}\left[\ket{\su\su\su\su} 
		+ \ket{\sd\sd\su\su} + \ldots + \ket{\sd\sd\sd\sd}\right] 
			\text{(even \# of }\su\text{)} \\
	\ket{\sqsd} &= \UP^\sd\ket{0} = 
		\dfrac{1}{2\sqrt{2}}\left[\ket{\sd\su\su\su} 
		+ \ket{\su\sd\su\su} + \ldots + \ket{\sd\sd\sd\su}\right] 
			\text{(odd \# of }\su\text{)}.
\end{align}

As expected,
\begin{align}
	\UP \ket{\sqsu} &= \UP \UP^\su\ket{0} = 
		\UP\invsqrt{2}(1 + \UP)\ket{0} = 
		\invsqrt{2}(\UP + 1)\ket{0} = \ket{\sqsu} \\
	\UP \ket{\sqsd} &= \UP \UP^\sd\ket{0} = 
		\UP\invsqrt{2}(1 - \UP)\ket{0} = 
		\invsqrt{2}(\UP - 1)\ket{0} = -\ket{\sqsd},
\end{align}
since $\UP^2 = 1$ for $Z_2$.

A physical state on a full lattice can be expressed as a linear combination of 
basis states which are products of operators $\UP^{\pm}$ for each plaquette:
\newcommand{\blambda}{\bm{\lambda}}%
\newcommand{\bchi}{\bm{\chi}}%
\begin{align}
	\label{eq:full-state}
	\ket{e_{\blambda}} &= \prod_{x} \UP^{\lambda_x}(x)\ket{0} \\
	\ket{\Psi} &= \sum_{\blambda} c_{\blambda} \ket{e_{\blambda}},
\end{align}
where $\blambda = (\lambda_x|\lambda_x \in \{\sd,\su\})$ 
are $M$-dimensional vectors that index the basis vectors of the physical states, 
$M$ is the number of plaquettes, and $c_{\blambda}$ are coefficients satisfying 
$\sum c_{\blambda}^2 = 1$. The basis states are orthonormal:
\begin{equation}
	\label{eq:basis-states-on}
	\braket{e_{\blambda}|e_{\blambda'}} = \bra{0}\prod_x \UP^{\lambda_x}(x) 
		\UP^{\lambda_x'}(x) \ket{0} = \prod_x \delta_{\lambda_x,\lambda_x'}
\end{equation}
and therefore we can write:
\begin{align}
	\label{eq:state-coeff}
 	c_{\blambda} = \braket{e_{\blambda}|\Psi}.
\end{align}

The orthogonality is apparent for two reasons. First, any product of the form 
$\UP^\su(x) \UP^\sd(x) = (1 - \UP^2(x))/2$ is zero, since $\UP^2(x) = 1$. 
Therefore, $\lambda_x$ must equal $\lambda_x'$ for all $x$ in order to get a 
non-zero result. Second, if all $\lambda_x = \lambda_x'$, then the left hand 
side of~\ref{eq:basis-states-on} reduces to:
\begin{equation}
	\bra{0}\prod_x (1 \pm \UP(x))\ket{0} = \braket{0|0} + \bra{0}
		\sum f(\UP(x))\ket{0},
\end{equation}
where $f(\UP(x))$ are various terms that contain at least one link operator. 
Such terms vanish, since they are anti-symmetric with respect to $\ket{0}$. In 
order to obtain a reduced density matrix, we can divide the set of plaquettes 
into two regions, $A$ and $\bar{A}$ and write
$\blambda= (\chi_x, \bar{\chi}_{\bar{x}}'|x \in A, \bar{x} \in \bar{A}) = 
\bchi \oplus \bar{\bchi}$ such that $\lambda_x = \chi_x$ if $x \in A$ and 
$\lambda_{\bar{x}} = \bar{\chi}_{\bar{x}}$ if $\bar{x} \in \bar{A}$. We can 
then write
\begin{align}
	\label{eq:prod-state}
	\ket{\Psi} &= \sum_{\bchi, \bar{\bchi}} c_{\bchi \otimes \bar{\bchi}} 
		\ket{e_{\bchi \otimes \bar{\bchi}}} \equiv \sum_{\bchi, \bar{\bchi}} 
			c_{\bchi, \bar{\bchi}} 
		\ket{e_{\bchi, \bar{\bchi}}}
\end{align}

The density matrix is then:
\begin{align}
	\label{eq:prod-rho}
	\rho[\blambda; \blambda'] = 
		\rho[\bchi,\, \bar{\bchi};\, \bchi',\,\bar{\bchi}'] = 
		c_{\bchi,\bar{\bchi}} c_{\bchi',\bar{\bchi}'}.
\end{align}

Consequently, the resulting reduced density matrix, $\rho_A$, is:
\begin{align}
	\label{eq:red-rho}
	\rho_A[\bchi; \bchi'] = \sum_{\bar{\bchi}} 
		c_{\bchi, \bar{\bchi}} c_{\bchi', \bar{\bchi}}
\end{align}

The entanglement entropy~\cite{Srednicki:1993im} is then:
\begin{equation}
	S_A = \tr\rho_A \ln\rho_A.
\end{equation}

The above representation of states is gauge invariant and the density matrix
$\rho$ is written explicitly in terms of vectors in the gauge invariant 
subspace. It follows that $S_A$ is gauge invariant.

One can also consider the transverse (or ``electric'') basis, which is the 
basis in which link operators, $L$, are diagonal:
\begin{align}
	L\ket{l} = l\ket{l}.
\end{align}

Specifically, in terms of link basis vectors:
\begin{align}
	L\ket{+} &= L\ket{\su} + L\ket{\sd} = \ket{\sd} + \ket{\su} = \ket{+} \\
	L\ket{-} &= L\ket{\su} - L\ket{\sd} = \ket{\sd} - \ket{\su} = -\ket{-}.
\end{align}

This basis is particularly useful due to its convenient and suggestive 
diagrammatic representation, which we will employ later. A gauge transformation 
at a lattice vertex $v$ is a product of link operators connected to that vertex 
and acting on a four link state as follows:
\newcommand{\hx}{\hat{x}}%
\newcommand{\hy}{\hat{y}}%
\begin{align}
	L_{v,v+\hx} L_{v,v+\hy} L_{v-\hx,v} L_{v-\hy,v} 
	\ket{l_{v,v+\hx} l_{v,v+\hy} l_{v-\hx,v} l_{v-\hy,v}} = 
		\nonumber\\
		l_{v,v+\hx} l_{v,v+\hy} l_{v-\hx,v} l_{v-\hy,v} 
		\ket{l_{v,v+\hx} l_{v,v+\hy} l_{v-\hx,v} l_{v-\hy,v}}.
\end{align}

Gauge invariant states must, therefore, satisfy 
$l_{v,v+\hx} l_{v,v+\hy} l_{v-\hx,v} l_{v-\hy,v} = 1$. This can 
be interpreted as a conservation law that ensures that an even number of 
$\ket{-}$ links are connected to every vertex. It can be seen that the 
resulting gauge invariant states take the form of linear combinations of closed 
loops of links in the $\ket{-}$ state. States can be manipulated using 
plaquette operators starting from $\ket{0}$, which, in the electric basis, is
equal to $\bigotimes \ket{+}$:
\newcommand{\bp}{\bm{p}}%
\begin{align}
	\label{eq:full-state-l}
	\ket{e_{\bp}} &= \prod_{x} (\UP(x))^{p_x}\ket{0} \\
	\ket{\Psi} &= \sum_{\bp} c_{\bp} \ket{e_{\bp}},
\end{align}
where $\bp = (p_x|p_x \in \{0,1\})$. Unsurprisingly, we recover the same 
structure as before (see Eq.~(\ref{eq:full-state})) and, consequently, the same 
dimensionality for the physical Hilbert space.

From a physical standpoint, the Hilbert space discussed in this section can 
either be seen as the space of closed electric loops or the space of magnetic 
fluxes going through plaquettes. From the magnetic perspective, it would seem 
natural that fluxes through extended areas are equal to the sum of individual 
fluxes through the elementary geometric constructs covered by that area.

\section{\label{sec:algebra}The algebra of gauge invariant operators in 2-d 
$Z_2$ gauge lattices}

There exists a duality between 2-d gauge theories and spin 
chains~\cite{PhysRevD.94.085029}, which was identified initially by Frank 
Wegner in~\cite{wegner}. We will summarize the relevant parts here. 

As seen previously, Wilson loops are gauge invariant. An algebra
is generated by the plaquette operators, $\UP(x)$. This completes the algebra
of gauge invariant operators that can be generated exclusively from link 
variables $U_{\mu}(x)$. The remaining gauge invariant operators are derived
from the link-flip operators $L_{\mu}(x)$. From the commutation 
relations in Eq.~(\ref{eq:link-op-comm}), we see that all $L_{\mu}(x)$ commute 
with each other. Since gauge transformations are a subalgebra of the algebra 
generated by link-flip operators, it follows that all $L_{\mu}(x)$ commute with 
gauge transformations and are, therefore, gauge invariant. Consequently, the 
full algebra of gauge invariant operators is generated by all $\UP(x)$ and all 
$L_{\mu}(x)$. Returning to the duality, the correspondence is:

\setlength{\tabcolsep}{18pt}
\begin{center}
\begin{tabular}{| c | c | c |}
	\hline
	Gauge theory & Spin chain & Comments\\
	\hline
	plaquette & spin & degrees of freedom\\
	\hline
	$\UP(x)$ & $\sigma_z(x)$ & - \\
	\hline
	$L_{\mu}(x + \hat{\mu}_\perp)$ & 
		$\sigma_x(x)\sigma_x(x + \hat{\mu}_\perp)$ & 
		$L_{\mu}(x)$ not at the edge, $\hat{\mu}\cdot\hat{\mu}_\perp = 0$ \\
	\hline
	$L^{\text{edge}}_{\mu}(x)$ & $\sigma_x(x)$ & 
		$L^{\text{edge}}_{\mu}(x)$ at the edge of the lattice\\
	\hline
\end{tabular}
\end{center}

\begin{figure}[tbp]
\centering
\reflectbox{\includegraphics[width=.50\textwidth]{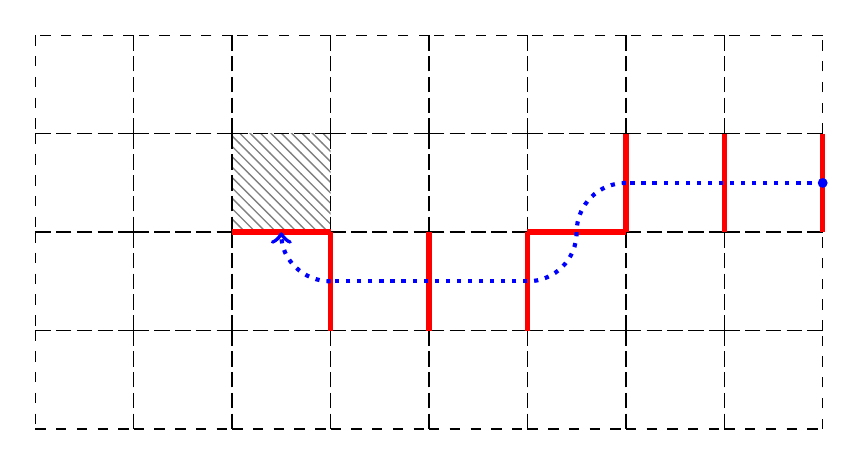}}
\caption{\label{fig:worm} Construction of one of the possible plaquette-flip 
operators. The operator acts on the shaded plaquette and is composed of 
$L_{\mu}(x)$ operators acting on links that are shown in thick, red, lines.}
\end{figure}

\newcommand{\LP}{L_{\square}}
There is no immediately obvious equivalent between $L_{\mu}(x)$ operators and 
the bulk $\sigma_x(x)$ operators. They can be \deleted{can be }constructed by 
observing that:
\begin{align}
	\sigma_x(x + \hat{\mu}_\perp) = 
		\sigma_x(x)\sigma_x(x)\sigma_x(x + \hat{\mu}_\perp) = 
		L_\mu^{\text{edge}}(x) L_\mu(x + \hat{\mu}_\perp),
\end{align}
where $\hat{\mu}\cdot\hat{\mu}_\perp = 0$. This can be generalized for 
arbitrary plaquettes (see FIG.~\ref{fig:worm}):
\begin{align}
	\LP(x) \equiv \sigma_x(x) = 
		L^{\text{edge}}_{\mu_0}(x_0) L_{\mu_1}(x_0 + \hat{\mu_0}_\perp)
		L_{\mu_2}(x_0 + \hat{\mu_0}_\perp + \hat{\mu_1}_\perp) \cdots 
		L_{\mu_n}(x).
\end{align}

Given a region $A$, the algebra generated by the operators 
$\UP(x), \LP(x)$ for $x \in A$ is a factor (an algebra with a trivial 
center).

\begin{figure}[tbp]
\centering
\reflectbox{\includegraphics[width=.28\textwidth]{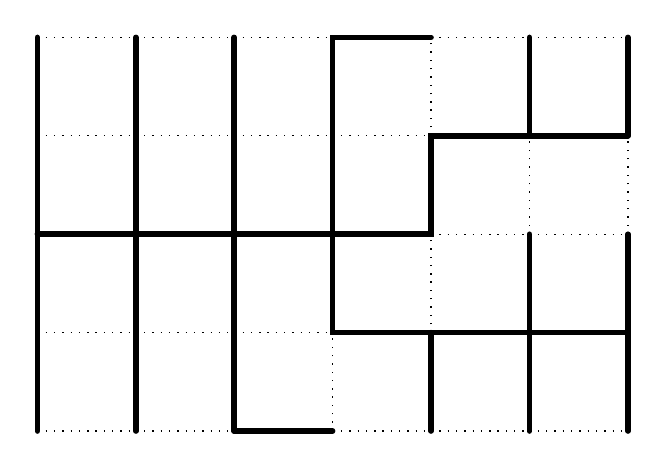}}
\caption{\label{fig:maximal-tree-gf} \added{Example of maximal tree gauge 
fixing in a simple gauge lattice. The thick links are fixed.}}
\end{figure}

Furthermore, the operators $\UP(x)$ and $\LP(x)$ are invariant under maximal
tree gauge fixing. This type of gauge fixing involves setting a certain set of 
links to a fixed value and only considering the remaining ones dynamic. This 
can be done as long as the fixed links do not form any loops (see, 
e.g.,~\cite{creutz1983quarks}\added{; also 
Figure~\ref{fig:maximal-tree-gf}}). The fixed links are typically 
set to the state $\ket{\su} = \ket{+} + \ket{-}$ and the corresponding 
$L_\mu(x)$ operator is removed from the algebra, since its inclusion would be 
at odds with the link being fixed. Because no loops of fixed links are allowed, 
we are guaranteed to have at least one dynamic edge link and we are also 
guaranteed that a path such as the one shown in FIG.~\ref{fig:worm} exists 
between a dynamic edge link and every plaquette. This, in turn, implies that 
all $L_\square(x)$ operators will exist in the algebra, but no particular 
$L_\mu(x)$ operator is guaranteed to be there.

We can also check what happens to the Hamiltonian from 
Eq.~(\ref{eq:hamiltonian}):
\begin{align}
	\label{eq:h-2d}
	H = \sum_{x, \mu} L_\mu(x) - \lambda\sum_x \UP(x) = 
		\sum_{\text{n.n.}} \LP(x)\LP(y) + \sum_{\text{edge}}\LP(x) -
		\lambda\sum_x \UP(x),
\end{align}
where the sums over $x$ are to be understood as sums over all links or all
plaquettes, respectively. The dual Hamiltonian is local, gauge invariant, and 
well defined for all choices of maximal tree gauge fixing conditions. With the
exception of the edge terms, the dual is a quantum transverse Ising model. This
dual represents the solution with $C=1$ in Eq.~(\ref{eq:ndof1}). 
The duality also underlines the problem with the inseparability proof 
in~\cite{buividovich_entanglement_2008}. If the observable theory consisted of 
an unconstrained spin network with Hilbert space $\Hs_0$, we would have no 
problem constructing a geometrical bipartition of the spin degrees of freedom. 
However, we can also construct a ``reverse'' Wegner dual gauge theory with a 
Hilbert space $\Hst$. On $\Hst$ there would be no bipartition of links 
supporting states in $\Hs_0$, but we should not use that to conclude that 
$\Hs_0$ is geometrically inseparable.

The idea floated previously of links being removed from the algebra under 
maximal tree gauge fixing deserves some more attention. If we adopt a temporal
gauge and also gauge-fix a particular time slice $t_0$ using a maximal tree, we 
are generally prevented from also fixing links in any subsequent time slice
\cite{creutzHamiltonianLattice}. The Hamiltonian will necessarily contain all 
link operators in the kinetic term. The link operators are then objects that 
relate link states at $t_0$ with link states at other times and a gauge fixing 
at $t_0$ remains associated with the absence of the ability to modify the state 
of certain links at $t_0$. Without a temporal gauge, one is free to use the 
exact same maximal tree of fixed links at all time slices. We can, therefore, 
completely remove the terms involving non-dynamical links from the Hamiltonian.
This issue is entirely hidden in the dual Hamiltonian in Eq.~(\ref{eq:h-2d}).

An alternative treatment to the entanglement entropy of Abelian lattice gauge 
theory based on the duality to spin systems can be found 
in~\cite{Radicevic:2016tlt}.

\section{\label{sec:two-plaq-lat}The two-plaquette lattice}

\begin{figure}[tbp]
\centering
\includegraphics[width=.30\textwidth]{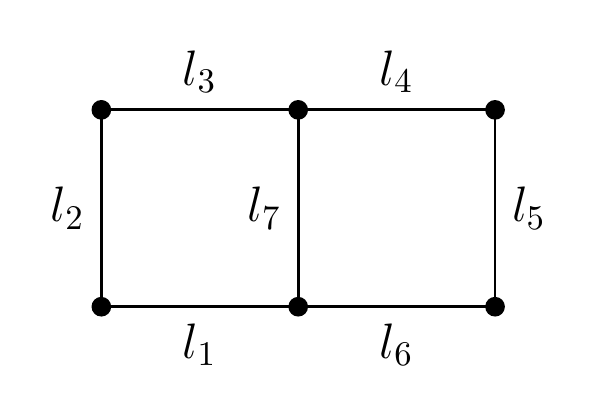}
\caption{\label{fig:simple-lattice} A simple two-plaquette lattice}
\end{figure}

The simplest two-dimensional pure gauge lattice setup is a two-plaquette
$Z_2$ lattice (FIG.~\ref{fig:simple-lattice}). It will be used to illustrate
some of the issues presented in the previous Section. 

The basic gauge invariant operators are:

\begin{align}
	&\UP^L = U_1 U_2 U_3 U_7 \\
	&\UP^R = U_4 U_5 U_6 U_7 \\
	&L_i, i \in \{1,..., 7\}.
\end{align}

The induced constraints are:

\begin{align}
	L_1 &= L_2 = L_3 \equiv \LP^L\\
	L_4 &= L_5 = L_6 \equiv \LP^R\\
	L_7 &= L_2 L_5 = \LP^L \LP^R.
\end{align}

The remaining gauge invariant operators can be obtained from $\UP^{L, R}$ 
and $\LP^{L, R}$. In particular, we can define:

\begin{align}
	U_\square^{\su\{L, R\}} &= 
		\dfrac{1}{\sqrt{2}} \left(1 + \UP^{\{L, R\}}\right) \\
	U_\square^{\sd\{L, R\}} &= 
		\dfrac{1}{\sqrt{2}} \left(1 - \UP^{\{L, R\}}\right),
\end{align}
which can be used to construct the magnetic basis:

\begin{align}
	\ket{\sqsd\sqsd} &= \UP^{\sd L} \UP^{\sd R} \ket{0} \\
	\ket{\sqsd\sqsu} &= \UP^{\sd L} \UP^{\su R} \ket{0} \\
	\ket{\sqsu\sqsd} &= \UP^{\su L} \UP^{\sd R} \ket{0} \\
	\ket{\sqsu\sqsu} &= \UP^{\su L} \UP^{\su R} \ket{0}.
\end{align}

\def\plht{2.2ex}
\newcommand{\plaqOOOO}
	{\vcenter{\hbox{\includegraphics[height=\plht]{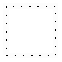}}}}
\newcommand{\plaqIIII}
	{\vcenter{\hbox{\includegraphics[height=\plht]{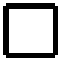}}}}
\newcommand{\plaqIIIO}
	{\vcenter{\hbox{\includegraphics[height=\plht]{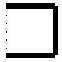}}}}
\newcommand{\plaqIOII}
	{\vcenter{\hbox{\includegraphics[height=\plht]{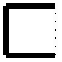}}}}

\newcommand{\dplaqOOOOOOO}
	{\vcenter{\hbox{\includegraphics[height=\plht]{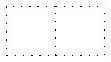}}}}
\newcommand{\dplaqOOOIIII}
	{\vcenter{\hbox{\includegraphics[height=\plht]{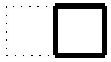}}}}
\newcommand{\dplaqIIIIOOO}
	{\vcenter{\hbox{\includegraphics[height=\plht]{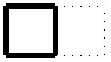}}}}
\newcommand{\dplaqIIIOIII}
	{\vcenter{\hbox{\includegraphics[height=\plht]{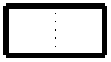}}}}
\newcommand{\dplaqOOIIOOO}
	{\vcenter{\hbox{\includegraphics[height=\plht]{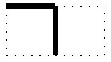}}}}
\newcommand{\dplaqOOOIOOI}
	{\vcenter{\hbox{\includegraphics[height=\plht]{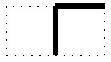}}}}
\newcommand{\dplaqOOIOOOI}
	{\vcenter{\hbox{\includegraphics[height=\plht]{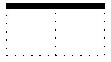}}}}

If we switch to the electric basis, we can express the tensor product structure
in a less abstract fashion through diagrams in which links in the $\ket{-}$ 
state are emphasized. The single plaquette states are 
$\ket{+}_\square = \ket{\plaqOOOO}, \ket{-}_\square = \ket{\plaqIIII}$, while 
the two-plaquette states are $\ket{++}_\square = \ket{\dplaqOOOOOOO}, 
\ket{-+}_\square = \ket{\dplaqIIIIOOO}, \ket{+-}_\square = \ket{\dplaqOOOIIII}, 
\ket{--}_\square = \ket{\dplaqIIIOIII}$. It is probably noteworthy that 
$\ket{\dplaqIIIOIII} = \ket{\plaqIIII} \otimes \ket{\plaqIIII} \ne 
\ket{\plaqIOII} \otimes \ket{\plaqIIIO}$. This is because the only gauge 
invariant operators that can change electric states are operators in the 
algebra of Wilson loops. If we start with $\ket{\dplaqOOOOOOO} = 
\ket{\plaqOOOO} \otimes \ket{\plaqOOOO}$ then $\ket{\dplaqIIIOIII} = 
U_\square^L U_\square^R\ket{\dplaqOOOOOOO} = \UP^L \UP^R 
\ket{\plaqOOOO} \otimes \ket{\plaqOOOO} = \ket{\plaqIIII} \otimes 
\ket{\plaqIIII}$.

\begin{figure}[tbp]
\centering
\includegraphics[width=.24\textwidth]{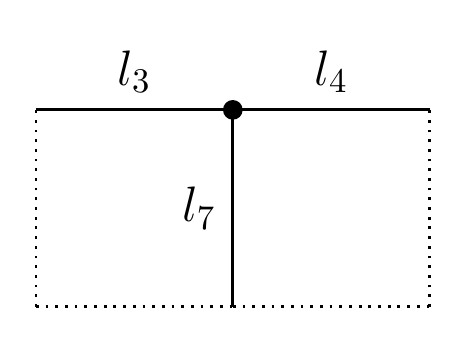}
\caption{\label{fig:simple-lattice-gf} Partial gauge fixing on the simple 
	lattice. The dotted links are set to $1$.}
\end{figure}

As pointed out in~\cite{Casini:2013rba}, certain sub-algebras are not factors.
In particular, if we chose to divide links into regions and considered the 
algebras of gauge invariant operators that can be constructed from the 
operators acting on links in each region, we would encounter algebras such
as $\Al_1 = \{\Id, L_1, L_2, L_3\}$ which satisfies $\Al_1 \cap (\Al_1)' = 
\Al_1 \ne \Id$. Alternatively the algebra of links including the entire left
loop would be $\Al_2 = \{\Id, L_1, L_2, L_3, L_7, \UP^L\}$, for which we would
find that $\{L_4, L_5, L_6\} \subset (\Al_2)''$, so $\Al_2$ also generates the
electric operators in the right loop and $(\Al_2)' \cap (\Al_2)'' \ne \Id$. 
Neither examples are specific to gauge theories. In a two spin system, 
$\Al_1 = \{\Id, \sigma_x^1\}$ also satisfies $\Al_1 \cap (\Al_1)' = 
\Al_1$, whereas for $\Al_2 = \{\Id, \sigma_x^1, \sigma_x^1\sigma_x^2, 
\sigma_z^1\}$ we would necessarily find that $\sigma_x^2 \in (\Al_2)' \cap 
(\Al_2)''$.

It may be interesting to compare the entanglement entropy in the magnetic, 
electric and kinematic spaces. We can simplify the lattice further by doing a 
partial gauge fixing (see FIG.~\ref{fig:simple-lattice-gf}). There is a 
single gauge transformation remaining, which, in the magnetic basis, flips all 
the remaining free links. Analyzing an arbitrary state can be somewhat 
unpalatable, so we stick to states of the form $\ket{\Psi} = 
\alpha\ket{\sqsd\sqsd} + \beta\ket{\sqsu\sqsu}$, with $\alpha^2 + \beta^2 = 1$. 
The entanglement entropy is:

\begin{align}
	S_A^{\text{mag.}} = -\alpha^2 \log \alpha^2 - \beta^2 \log \beta^2
\end{align}
and it can vary between zero (for $\alpha = 0$ or $\beta = 0$) and $\log 2$ 
(for $\alpha = \beta = 0.5$). The kinematic state is:

\begin{equation}
	\ket{\Psi} = 
		\dfrac{\alpha}{\sqrt{2}}\left[\ket{\su_3\sd_7\su_4} + 
			\ket{\sd_3\su_7\sd_4}\right] +
		\dfrac{\beta}{\sqrt{2}}\left[\ket{\su_3\su_7\su_4} + 
			\ket{\sd_3\sd_7\sd_4}\right].
\end{equation}

The reduced density matrix obtained by tracing over links $7$ and $4$ is:

\begin{align}
	\rho_{7,4} &= \dfrac{\alpha^2}{2} \ket{\su_3}\bra{\su_3} + 
				  \dfrac{\alpha^2}{2} \ket{\sd_3}\bra{\sd_3} + 
				  \dfrac{\beta^2}{2} \ket{\su_3}\bra{\su_3} + 
				  \dfrac{\beta^2}{2} \ket{\sd_3}\bra{\sd_3} \\
			   &= \dfrac{1}{2}\ket{\su_3}\bra{\su_3} + 
			   	  \dfrac{1}{2} \ket{\sd_3}\bra{\sd_3}.
\end{align}

The resulting entanglement entropy is now $S_A^{\text{kin.}} = \log 2$, 
independent of $\alpha$ and $\beta$. This value is not gauge invariant since
fixing, e.g., link $7$ to $\ket{\su_7}$ yields 
$\ket{\Psi} = \alpha \ket{\sd_3\su_7\sd_4} + \beta \ket{\su_3\su_7\su_4}$, and 
we recover the value of $S_A^{\text{mag.}}$.

In the electric basis, after the gauge fixing employed above, the basis vectors 
are $\ket{++}_\square = \ket{\dplaqOOOOOOO} = \ket{+_3 +_7 +_4},\,
\ket{-+}_\square = \ket{\dplaqOOIIOOO} = \ket{-_3 -_7 +_4},\, \ket{+-}_\square 
= \ket{\dplaqOOOIOOI} = \ket{+_3 -_7 -_4},\, \ket{--} = \ket{\dplaqOOIOOOI} = 
\ket{-_3 +_7 -_4}$. The state $\ket{\Psi}$ is:

\begin{align}
	\ket{\Psi} = 
		\dfrac{\alpha + \beta}{2}
			\left[\ket{+_3 +_7 +_4} + \ket{-_3 +_7 -_4}\right] -
		\dfrac{(\alpha - \beta)}{2}
			\left[\ket{+_3 -_7 -_4} + \ket{-_3 -_7 +_4}\right].
\end{align}

The electric vectors correspond to states with electric fluxes conserved at
the shared lattice site. The state expressed in the electric basis has the same 
form as the kinematic state and one can conclude that $S_A^{\text{elec.}} = 
\log 2$. In the electric picture, one would attribute the entanglement entropy 
to the constraints on the electric degrees of freedom. Once again, if we were 
to fully fix the gauge and remove one of the links $l_i$ from the state, the 
entanglement entropy of the projected state would take the same value as 
$S_A^{\text{mag.}}$. The exact choice of $i$ is not relevant for the state in 
this example.

\section{\label{sec:pbc}Periodic boundary conditions}

The introduction of periodic boundary conditions in the spatial direction is
associated with two new phenomena. One is:

\begin{align}
	\label{eq:global-plaq-constraint}
	\prod_x U_\square(x) = 1, 
\end{align}
since every link operator $U_\mu(x)$ appears exactly twice in the product. The
second is the addition of two new independent topological degrees of freedom,
associated with loops that wind around the two spatial dimensions.

The global constraint in Eq.~(\ref{eq:global-plaq-constraint}) is a 
manifestation of the magnetic Gauss' law, which holds that the net magnetic 
flux through a closed surface is zero. Once again, we can consider the 
possibility of an unconstrained duality. In the spirit of Eq.~(\ref{eq:ndof1}), 
for a 2-d lattice with periodic boundary conditions:
\newcommand{\Nplaq}{N_{\text{plaq.}}}%
\begin{align}
	\Ndof^d &= (C + 4P) \Nplaq^d + V N_{\text{vert.}}^d + 
		L N_{\text{links}}^d.
\end{align}

We can notice that in a lattice with periodic boundary conditions in both $x$ 
and $y$ directions we have $N_{\text{links}}^d = 2\Nplaq^d = 
2N_{\text{vert.}}^d$. We can then write:
\begin{align}
	\label{eq:dofs-pbc}
	\Ndof^d = (C + 4P + V + 2L)\Nplaq^d.
\end{align}

To see that there is no \replaced{homogeneous and 4-fold isotropic 
unconstrained dual lattice with periodic boundary conditions that can support
a suitable duality, we can start with a lattice with dimensions 
$N_x = 2^p, N_y = 2^q$. Without the global constraint, there would be precisely 
one physical degree of freedom per plaquette. However, the presence of the 
global constraint reduces the number of physical degrees of freedom by one, 
leading to }{unconstrained dual in general, we can start with a 
lattice with dimensions $N_x = 2^p, N_y = 2^q$ such that }
$\Ndof = N_x N_y - 1 = 2^{p + q} - 1$. $\Ndof$ is a Mersenne number and some 
$p$ and $q$ would lead to $\Ndof$ being prime\deleted{, leaving us with either 
$N_x N_y = 1$ or $N_x = 1$ or $N_y = 1$, which represent zero and 
one-dimensional geometries}. \added{The only solutions to $\Ndof = \Ndof^d$ in 
Eq.~\ref{eq:dofs-pbc} are $\Nplaq^d = 1$ and $\Nplaq^d = N_x^d N_y^d = \Ndof$. 
The first solution corresponds to a zero-dimensional geometry, while the second 
implies either that $N_x^d = 1$ or that $N_y^d = 1$, which are one-dimensional 
geometries.}

The global constraint cannot, therefore, be eliminated without severely 
spoiling the geometry. However, using the plaquette basis, we can still work in 
a gauge invariant space. As before (see Eq.~(\ref{eq:full-state})), we express 
a state as a linear combination of basis vectors for some set $X$ of 
independent plaquettes:
\begin{align}
	\ket{e_{\blambda}} = \prod_{x \in X} \UP^{\lambda_x}(x) \ket{0}.
\end{align}

We can subsequently extract the coefficients of a gauge invariant state on the 
constrained space using products of $\UP$ operators over all plaquettes:
\newcommand{\bxi}{\mathbf{\xi}}
\begin{align}
	\braket{e_{\blambda,\lambda_0}|e_{\blambda'}} = 
		\bra{0}
		\UP^{\lambda_0}(x_0)
		\prod_{x \in X} \UP^{\lambda_x}(x) 
		\prod_{x' \in X} \UP^{\lambda_{x'}}(x')\ket{0},
\end{align}
where $x_0$ represents the remaining plaquette ($x_0 \notin X)$. Since 
$\UP^{\lambda_x}(x)$ satisfy $\UP^{\lambda_x}(x)\UP^{\lambda_x'}(x) = 
2\delta_{\lambda_x, \lambda_x'}\UP^{\lambda_x}(x)$ and we have:

\begin{align}
	\braket{e'_{\blambda',\lambda_0'}|e_{\blambda}} &= 
		\delta_{\blambda, \blambda'}
		\bra{0}
		\UP^{\lambda_0'}(x_0)
		\prod_{x \in X} \UP^{\lambda_x}(x) \ket{0} \nonumber\\
		 &= \dfrac{1}{2}\delta_{\blambda, \blambda'}
		\bra{0} 1 + \prod_x p(\lambda_x') \UP(x) + ...\ket{0} \nonumber\\
		&= \dfrac{1}{2}\delta_{\blambda, \blambda'} 
			\delta(1 + \prod_x p(\lambda_x'))
\end{align}
where $p(\su) = 1$ and $p(\sd) = -1$ and the ellipsis stands for terms that
are antisymmetric with respect to $\ket{0}$. When the constraint is satisfied,
there is an even number of plaquettes in the $\ket{\sd}$ state and the product
over $p(\lambda_x)$ is equal to $1$ leading to 
$\braket{e'_{\blambda',\bxi}|e_{\blambda}} = 1$. Conversely, when the 
constraint is not satisfied, the product over $p(\lambda_x)$ is $-1$ and the
dot product vanishes. The constrained space density matrix corresponding to
some state $\ket{\Psi}$ is then:
\begin{align}
	\rho[\blambda, \lambda_0; \blambda', \lambda_0'] = 
		\braket{\Psi|e_{\blambda',\lambda_0'}}
			\braket{e_{\blambda,\lambda_0}|\Psi}
\end{align}

One can then divide degrees of freedom using $\blambda \oplus \lambda_0 = 
\bchi \oplus \bar{\bchi}$ and proceed as in 
Eq.~(\ref{eq:prod-state})-(\ref{eq:red-rho}).

With no edge links in the original Hamiltonian (see Eq.~(\ref{eq:hamiltonian})), 
the plaquette basis Hamiltonian would take the form:
\begin{align}
	H = \sum_{x, \mu} L_\mu(x) - \lambda\sum_x \UP(x) = 
		\sum_{\text{n.n.}} \LP(x)\LP(y) -
		\lambda\sum_x \UP(x),
\end{align}
which is the transverse Ising model. In the $\lambda \rightarrow 0$ limit
there is no magnetic term and the entanglement entropy is $S_A = \log 2$ and
topological. In the $\lambda \rightarrow \infty$ limit the plaquettes are 
polarized and $S_A = 0$.

\begin{figure}[tbp]
\centering
\includegraphics[width=.95\textwidth]{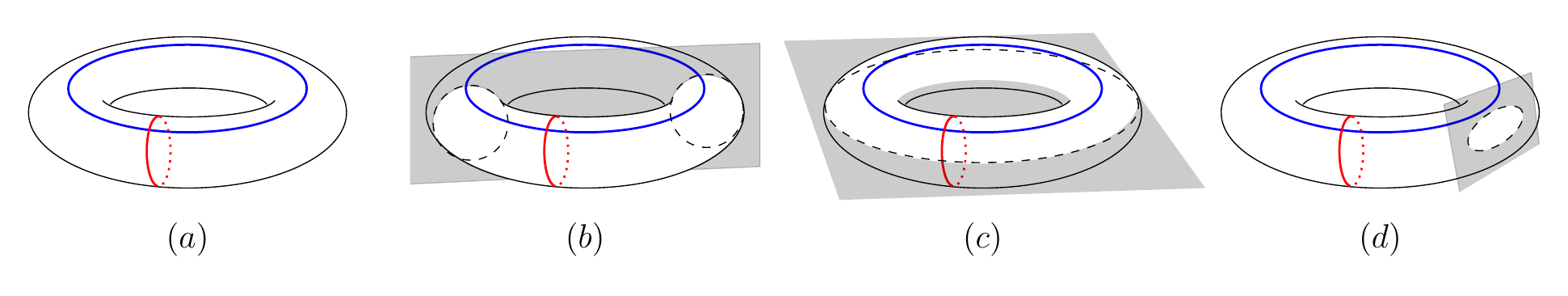}
\caption{\label{fig:torus-dof} Topological degrees of freedom on a torus (a) 
and various bipartition choices (b), (c), (d). Bipartition boundaries are
shown as dashed lines.}
\end{figure}

We now return to the topological degrees of freedom. The topology induced by 
imposing periodic boundary conditions on the two-dimensional lattice is that of 
a torus. The two non-plaquette degrees of freedom are related to the magnetic 
flux through the inside and center of the torus 
(see FIG.~\ref{fig:torus-dof}a). There is no preferential choice for the 
generating algebra of the topological degrees of freedom and the different 
choices can be related using plaquette operators. However, given such a choice, 
there exist two classes of bipartitions based on whether exactly one region 
contains both topological loops (FIG.~\ref{fig:torus-dof}d) or not 
(FIG.~\ref{fig:torus-dof}bc). In the former case, at least one of the 
topological loops cannot be expressed as a non-trivial tensor product. To the 
extent that one is comfortable with the idea of separating what appears to be 
an atomic degree of freedom, the problem can be alleviated by enlarging the 
Hilbert space (see, e.g.,~\cite{Donnelly:2011hn}) or, as will be shown in 
Section~\ref{sec:matter}, by coupling to matter fields.

One encounters a similar situation when imposing periodic boundary conditions
in a single direction leading to a cylindrical topology. The global constraint
is not present, but a topological degree of freedom remains. It is associated
with the magnetic flux through the cylinder.

\section{\label{sec:three-d}$3+1$ dimensions}

\begin{figure}[tbp]
\centering
\includegraphics[width=.15\textwidth]{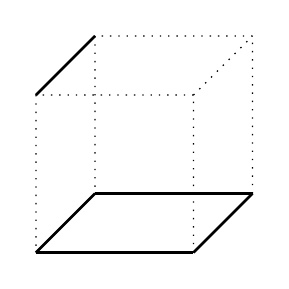}
\caption{\label{fig:cube-gf} Maximal tree gauge fixing on an elementary cube. 
The dotted links are fixed.}
\end{figure}

The three-dimensional case is characterized by the existence of a local 
magnetic Gauss constraint of a similar nature as the one in 
Eq.~(\ref{eq:global-plaq-constraint}):
\begin{align}
	\label{eq:constraint-3d}
	\prod_{C}U_{\square\mu\nu}(x) = 1, 
\end{align}
where $C$ represents the faces of an elementary cube. The constraint is not a 
gauge constraint since it cannot be eliminated by gauge fixing. For example, a 
maximal tree gauge fixing on a single cube would result in five remaining 
dynamic links as shown in FIG.~\ref{fig:cube-gf} and Eq.~\ref{eq:constraint-3d} 
would still hold.

\begin{figure}[tbp]
\centering
\includegraphics[width=.4\textwidth]{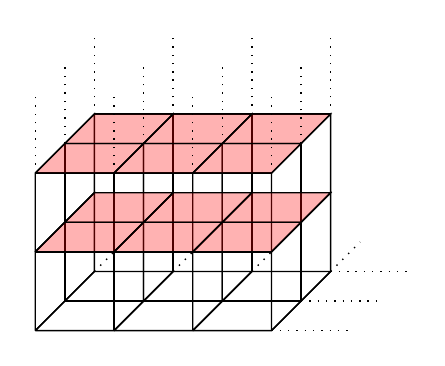}
\caption{\label{fig:maximal-manifold} Example of unconstrained Hilbert space in 
three-dimensional time slices. The degrees of freedom on the red/shaded 
plaquettes can be expressed in terms of other degrees of freedom.}
\end{figure}

To look at the space of unconstrained degrees of freedom we can find a subset 
of plaquettes such that no closed surfaces are formed, a procedure reminiscent 
of the maximal tree gauge fixing procedure. One such subset is shown in 
FIG.~\ref{fig:maximal-manifold}. A count of the number of degrees of freedom 
yields $N_{d.o.f.} = 2 N_x N_y N_z + N_x N_y + N_y N_z + N_z N_x$, where $N_i$
is the number of plaquettes in direction $i$. Consequently, a dual with 
unconstrained degrees of freedom would necessarily be both anisotropic and
would have non-local terms in the action. The density matrix can be written
in terms of the constrained but gauge-invariant degrees of freedom as it was
done in the previous section. The Hamiltonian takes the following form:

\begin{align}
	H = \sum_{x, \mu} \prod_{i \in ST(x, \mu)} \LP^i - \lambda\sum_{x} \UP(x),
\end{align}
where $ST(x, \mu)$ (the ``staple'') is the set of all plaquettes that contain 
the link $L_\mu(x)$.

As it was mentioned earlier, in a Wilsonian $2+1$ dimensional theory a time 
slice would include plaquettes with a time component, making it a restricted 
form of a three dimensional space in which no closed surfaces would exists, 
corresponding to only the bottom layer in FIG.~\ref{fig:maximal-manifold}. 
The resulting physical Hilbert space would remain unconstrained.

\section{\label{sec:matter}Coupling to matter fields}

Matter fields in lattice theories are associated with degrees of freedom that
live on the vertices of the lattice. They transform under a gauge 
transformation as:
\begin{align}
	\phi(x) \rightarrow g(x) \phi(x).
\end{align}

This implies a new set of gauge-invariant quantities:
\begin{align}
	\phi^\dagger(x) \phi(x) \rightarrow \phi^\dagger(x) 
		g^\dagger(x) g(x) \phi(x) = \phi^\dagger(x) \phi(x).
\end{align}

Additionally, since end points of gauge links transform in the same way as 
matter fields, we can also identify products of the following form as gauge 
invariant:

\begin{align}
	\phi^\dagger(x_1) &u_{\mu_1}(x_1) \ldots 
		u_{\mu_{n-1}}(x_{n-1})\phi(x_n) \rightarrow \nonumber\\
	&\phi^\dagger(x_1) g^\dagger(x_1) g(x_1) u_{\mu_1}(x_1) 
		g^\dagger(x_1 + \hat{\mu}_1) \ldots \nonumber\\
	&g(x_{n-2}) u_{\mu_{n-1}}(x_{n-1})g^\dagger(x_{n-1} + \hat{\mu}_{n-1})
			g(x_n)\phi(x_n),
	\label{eq:open-strings}
\end{align}
provided that $x_i + \hat{\mu}_i = x_{i+1}$. In other words, Wilson lines
multiplied with matter fields at the ends are gauge invariant quantities. We 
can employ a slight re-definition of the fields in order to separate degrees of
freedom into gauge-dependent and gauge-independent quantities:

\begin{align}
	\phi(x) = \left|\phi(x)\right|v(x) = \tilde{\phi}(x)v(x),
\end{align}

with $v(x)$ being gauge group valued. The fields $\tilde{\phi}(x)$ now 
transform trivially under a gauge transformation. Using the $v(x)$ fields, 
which inherit the transformation properties of $\phi(x)$, we can introduce 
gauge invariant degrees of freedom associated with individual links:
\begin{align}
	\tilde{u}_\mu(x) = v^\dagger(x) u_\mu(x) v(x + \hat{\mu}).
\end{align}

When a product of $\tilde{u}_\mu(x)$ variables is taken over a closed loop,
the $v(x)$ fields cancel and we obtain:
\begin{align}
	\prod_{(x, \mu) \in C} \tilde{u}_\mu(x) = 
		\prod_{(x, \mu) \in C} u_\mu(x).
\end{align}

The operators that are diagonalized by the vectors $\ket{v(x)}$ and 
$\ket{u(x)}$ generate part of the gauge invariant algebra through products of 
the following form:
\newcommand{\UT}{\tilde{U}}%
\newcommand{\tPhi}{\tilde{\Phi}}%
\begin{align}
	\UT_\mu(x) = V^\dagger(x) U_\mu(x) V(x + \hat{\mu}),
\end{align}
where $V(x)\ket{v(x)} = v(x)\ket{v(x)}$ and $U_\mu(x)\ket{u_\mu(x)} = 
u_\mu(x)\ket{u_\mu(x)}$. The remaining gauge invariant operators are $K(x)$ and
$L_\mu(x)$, which are conjugates to $V(x)$ and $U_\mu(x)$, respectively. 
Together with $\UT_\mu(x)$ and the algebra of the fields $\tPhi(x)$, 
they generate the full gauge invariant algebra.

\begin{figure}[tbp]
\centering
\includegraphics[width=.75\textwidth]{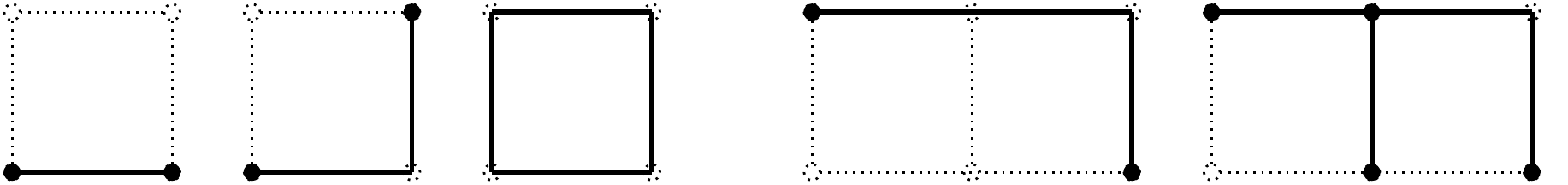}
\caption{\label{fig:example-matter-el-states} Example electric states on simple
matter and gauge lattices with a $Z_2$ gauge group. The dotted links/vertices 
are in the $\ket{+}$ state, while the thick/filled links/vertices are in the 
$\ket{-}$ state.}
\end{figure}

As we did in the case of pure gauge theories, we can explicitly see the form
of the states in the electric basis by applying products of $\UT_\mu(x)$ 
operators to the vacuum $\ket{0} = \sum_{v(x), u_\mu(x)} 
\bigotimes_x\ket{v(x)}\bigotimes_{x, \mu}\ket{u_\mu(x)}$. Specifically, for a 
$Z_2$ gauge group, the gauge invariant states can be visualized as the set of 
all superpositions of non-overlapping arbitrary length strings. The constraint
corresponding to gauge invariance dictates that each vertex must be in the 
$\ket{-}$ state iff there is an odd number of $\ket{-}$ links connected to it. 
This can be seen by looking at how a gauge transformation acts on the state of 
a vertex and the links connected to it in the electric basis:
\begin{align}
	G\ket{k(x_0)}\otimes \bigotimes_{x, \mu} \ket{l_\mu(x)} 
		&= K(x_0)\prod_{x, \mu} L_\mu(x)\ket{k(x_0)}\otimes 
			\bigotimes_{x, \mu} \ket{l_\mu(x)} \nonumber\\
		&= k(x_0) \prod_{x, \mu} l_\mu(x) \ket{k(x_0)}\otimes 
			\bigotimes_{x, \mu} \ket{l_\mu(x)},
\end{align}
where $\ket{k}$ and $\ket{l_i}$ are eigenstates of $K$ and $L_i$ with 
eigenvalues $k$ and $l_i$, respectively and $x, \mu$ represent links that have
one endpoint at $x_0$. The requirement that the gauge 
transformation be the identity for physical states implies that 
$k\prod_i l_i = 1$ or $k = \prod_i l_i$. Some example gauge invariant electric 
states in small two dimensional lattices are shown in 
FIG.~\ref{fig:example-matter-el-states}.

The unconstrained gauge part of the algebra, $\Al_g$ is generated by the 
$L_\mu(x)$ and $\UT_\mu(x)$ operators which act on the Hilbert space of gauge
invariant link states $\Hs_g$. To get the full unconstrained gauge invariant
algebra, one adds the algebra of the fields $\tPhi$, $\Al_{\tPhi}$. The choice
of degrees of freedom on the links and vertices corresponds to $L = 1, V = 1$ 
in Eq.~(\ref{eq:ndof1}). The commutation relations of the operators in $\Al_g$ 
can be inferred from the commutation relations of the non gauge invariant 
operators:
\begin{align}
	[L_\mu(x), L_\nu(y)] &= 0 \nonumber\\
	[\UT_\mu(x), \UT_\nu(y)] &= 0
			\nonumber \\
	[L_\mu(x), \UT_\nu(y)] &= 
		V^\dagger(y)\left[L_\mu(x), U_\nu(y)\right]V(y + \hat{\nu}).
\end{align}

The last commutator is zero if $[L_\mu(x), U_\nu(y)] = 0$, which is 
true if $x \ne y$ or $\mu \ne \nu$. As in the pure gauge case, the electric 
basis on the full lattice can be built using $\UT_\mu(x)$ operators and the
ground state satisfying $L_\mu(x)\ket{0} = \ket{0}, \forall x, \mu$:
\begin{align}
	\ket{e_{\blambda}} = \prod_{x, \mu} \UT_\mu^{\lambda_{x, \mu}}(x)\ket{0},
\end{align}
where $\lambda_{x, \mu} \in \{0, 1\}$. A basis for the full gauge invariant 
Hilbert space of the theory would then be formed by tensor products of vectors 
$\ket{e_{\blambda}}$ and basis vectors in the Hilbert space of the fields 
$\tPhi$, allowing us to write $\Hs_0 = \Hs_g \otimes \Hs_{\tPhi} = 
(\Hs_{g,A} \otimes \Hs_{\tPhi,A}) \otimes (\Hs_{g,B} \otimes \Hs_{\tPhi,B})$.

\section{\label{sec:surface-charges}Surface charges}

The case of a theory with surface charges is a special case of coupling to 
matter fields where matter fields are only defined on the boundaries of a 
lattice. We allow for both dynamic and non-dynamic surface charges, but
restrict ourselves to $Z_2$ gauges and two dimensions for simplicity. We can 
employ the basis used in the previous section where we decouple the gauge 
portion from the matter fields. In the electric basis, the gauge invariant 
states take the form of loops and strings that open on boundaries. We are 
concerned with whether strings can be expressed as tensor products of vectors 
in bipartitions of plaquettes. We use the diagrammatic representation of states 
since it provides a more clear picture.

There are two non-trivial situations: bipartitions in which both regions share 
some of the lattice boundary and bipartitions in which one region is entirely
in the bulk of the lattice. In the first case, we seek a tensor product for
strings that cross the boundary between regions once, while in the second, we 
are concerned with open strings that cross the bulk region. The two cases are 
not exhaustive, but illustrate the Hilbert space factorization where it is less
obvious.

Before proceeding, we note that all open strings along curves $C_b$ with links 
in the bulk can be created by acting on the vacuum with products of 
gauge invariant link operators on the edge $\UT_i, i \in C_e$ to create a 
closed curve with $C_b$ and then acting on the result with all plaquette 
operators in the surface enclosed by the curve $C_e \cup C_b$. For example:
\newcommand{\TL}{\tilde{L}}%
\newcommand{\Up}{\tilde{U}_\square}%
\newcommand{\bglElb}
	{\vcenter{\hbox{\includegraphics[height=7.0ex]{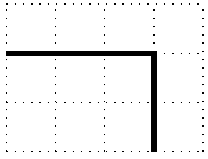}}}}%
\newcommand{\bglVacBlLabeled}
	{\vcenter{\hbox{\includegraphics[height=7.0ex]{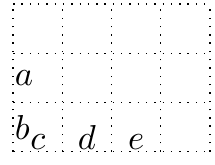}}}}%
\newcommand{\bglBlPlLabeled}
	{\vcenter{\hbox{\includegraphics[height=7.0ex]{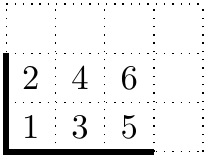}}}}%
\begin{align}
	\left|\bglElb\right> = 
		\Up(1)...\Up(6)	\left|\bglBlPlLabeled\right> = 
		\Up(1)...\Up(6)	\UT_a...\UT_e\left|\bglVacBlLabeled\right>.
\end{align}

For more clarity, we can express operators in a diagrammatic form:

\newcommand{\bglOpU}
	{\vcenter{\hbox{\includegraphics[height=7.0ex]{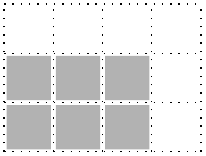}}}}%
\newcommand{\bglOpE}
	{\vcenter{\hbox{\includegraphics[height=7.0ex]{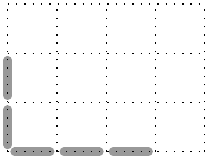}}}}%
\begin{align}
	\Up(1)...\Up(6) &= \bglOpU \\
	\UT_a...\UT_e &= \bglOpE,
\end{align}

leading to
\newcommand{\bglVacBl}
	{\vcenter{\hbox{\includegraphics[height=7.0ex]{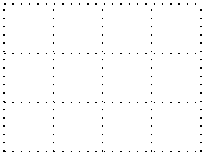}}}}%
\newcommand{\bglBlPl}
	{\vcenter{\hbox{\includegraphics[height=7.0ex]{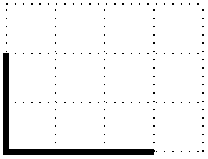}}}}%
\begin{align}
	\left|\bglElb\right> = 
		\bglOpU \left|\bglBlPl\right> = 
		\bglOpU	\bglOpE\left|\bglVacBl\right>.
\end{align}

The space of bulk strings exhibits a global symmetry due to the fact that one
can close bulk strings using edge strings in two ways. This symmetry is 
associated with the constraint $\prod_{i \in C_\text{edge}} \UT_i = \prod_x 
\Up(x)$ where $C_\text{edge}$ is the set of links on the edge of the lattice. 
The constraint can be re-written as $\prod_{i \in C_{e}^1} \UT_i 
\prod_{x \in V} \Up(x) = \prod_{i \in C_{e}^2} \UT_i \prod_{x \in \bar{V}} 
\Up(x)$, with $C_{e}^1 \cup C_{e}^2 = C_\text{edge}$. Symmetries prevent
us from removing this constraint. The tensor product structure arises naturally 
from the bulk plaquette algebra and the edge gauge invariant link algebra 
resulting in identities of the following form:
\newcommand{\bglS}
	{\vcenter{\hbox{\includegraphics[height=7.0ex]{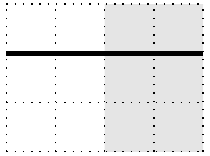}}}}%
\newcommand{\bglSA}
	{\vcenter{\hbox{\includegraphics[height=7.0ex]{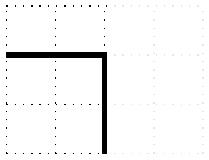}}}}%
\newcommand{\bglSB}
	{\vcenter{\hbox{\includegraphics[height=7.0ex]{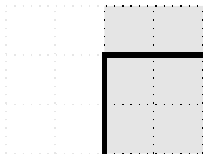}}}}%
\newcommand{\bglSAVac}
	{\vcenter{\hbox{\includegraphics[height=7.0ex]{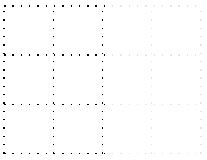}}}}%
\newcommand{\bglSBVac}
	{\vcenter{\hbox{\includegraphics[height=7.0ex]{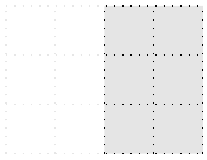}}}}%
\newcommand{\bglSAOpA}
	{\vcenter{\hbox{\includegraphics[height=7.0ex]{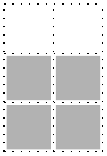}}}}%
\newcommand{\bglSAOpB}
	{\vcenter{\hbox{\includegraphics[height=7.0ex]{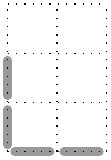}}}}%
\newcommand{\bglSAOpC}
	{\vcenter{\hbox{\includegraphics[height=7.0ex]{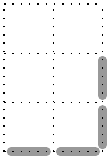}}}}%
\begin{align}
	\left|\bglS\right> &= \left|\bglSA\right> \otimes \left|\bglSB\right>,
\end{align}
with
\begin{align}
	\left|\bglSA\right> &= 
		\bglSAOpA_A \cdot \bglSAOpB_A \left|\bglSAVac\right>, \\
	\left|\bglSB\right> &=
		\bglSAOpA_B \cdot \bglSAOpC_B \left|\bglSBVac\right>,
\end{align}
and
\newcommand{\hgtb}{8.0ex}
\newcommand{\bglQ}
	{\vcenter{\hbox{\includegraphics[height=\hgtb]{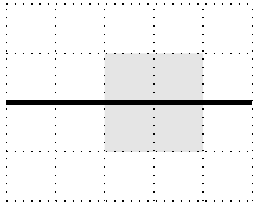}}}}%
\newcommand{\bglQA}
	{\vcenter{\hbox{\includegraphics[height=\hgtb]{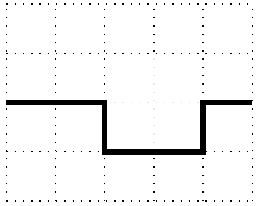}}}}%
\newcommand{\bglQB}
	{\vcenter{\hbox{\includegraphics[height=\hgtb]{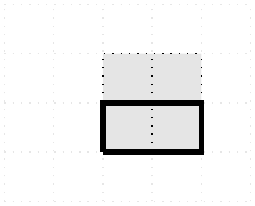}}}}%
\newcommand{\bglQAVac}
	{\vcenter{\hbox{\includegraphics[height=\hgtb]{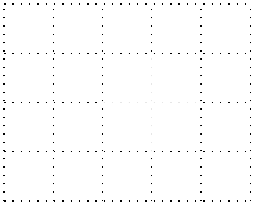}}}}%
\newcommand{\bglQBVac}
	{\vcenter{\hbox{\includegraphics[height=\hgtb]{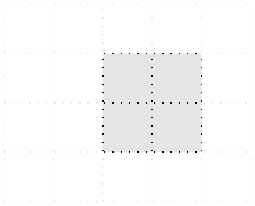}}}}%
\newcommand{\bglQAOpA}
	{\vcenter{\hbox{\includegraphics[height=\hgtb]{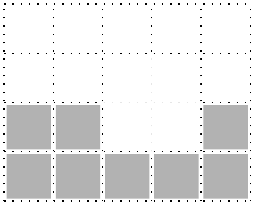}}}}%
\newcommand{\bglQAOpB}
	{\vcenter{\hbox{\includegraphics[height=\hgtb]{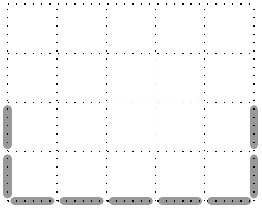}}}}%
\newcommand{\bglQBOpC}
	{\vcenter{\hbox{\includegraphics[height=\hgtb]{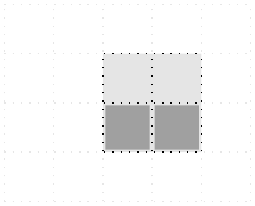}}}}%
\begin{align}
	\left|\bglQ\right> &= \left|\bglQA\right> \otimes \left|\bglQB\right>,
\end{align}
with
\begin{align}
	\left|\bglQA\right> &= \bglQAOpA_A\bglQAOpB_A\left|\bglQAVac\right>, \\
	\left|\bglQB\right> &= \bglQBOpC_B\left|\bglQBVac\right>.
\end{align}

\section{Concluding remarks}

The use of the $Z_2$ group throughout the paper was motivated by its simplicity
and the fact that it enables a useful diagrammatic notation for states. We 
expect the extension to other Abelian groups to be relatively straightforward. 
In a $U(1)$ theory in the link basis $\ket{e^{i\phi}}$ states take the form of 
periodic functions, whereas electric states are discrete 
(see e.g.,~\cite{u1lattice}) and represented by the Fourier modes of the link
basis states. The gauge invariant algebra is generated by loop operators 
$\UP(x), \UP^*(x),$ electric operators $L = - i \partial/\partial_\phi$ with
electric eigenstates $\ket{q} = \int e^{iq\phi}\ket{e^{i\phi}}$ such that
$L\ket{q} = q\ket{q}$, and link rotation operators $L^\theta\ket{e^{i\phi}} = 
\ket{e^{i(\phi + \theta)}}$ such that $L^\theta\ket{q} = e^{i\theta q}\ket{q}$. 
The requirement of gauge invariance at a vertex in the electric basis then 
reads:
\begin{align}
	\ket{q_1 q_2 q_3 q_4} &= 
		G\ket{q_1 q_2 q_3 q_4} \nonumber\\
	&= L_1^\theta L_2^\theta L_3^{-\theta} L_4^{-\theta}
		\ket{q_1 q_2 q_3 q_4} \nonumber\\ 
    &= e^{i\theta(q_1 + q_2 - q_3 - q_4)} 
    	\ket{q_1 q_2 q_3 q_4},
\end{align}
which must be satisfied for all $\theta$, implying $q_1 + q_2 - q_3 - q_4 = 0$. 
In other words, the electric fluxes are conserved at vertices. Similar to the 
$Z_2$ theory, such states can be created by acting on the vacuum with no 
electric fluxes with operators $\UP(x)$ which raise the electric flux around a 
plaquette by one and $\UP^*(x)$ which lower it.

The case of $3+1$ dimensional theories with periodic boundary conditions is 
absent. It is a straightforward extension of sections \ref{sec:pbc} and
\ref{sec:three-d}. Similarly, the Wilsonian $3+1$ dimensional theory was left 
out.

The analysis performed in this paper is only valid for discretized spaces. 
As shown in~\cite{Callan:1994py}, the entanglement entropy is UV divergent. 
Furthermore, as Witten argues in~\cite{Witten:2018zxz}, Hilbert spaces 
supported on geometries dense in some connected space may not be separable 
precisely because the UV divergence of the entanglement entropy is a universal 
feature not tied to a particular state.

\section{Acknowledgements}
This work is supported in part by the US Department of Energy grant DE-SC-000999.

\bibliography{ee}
\end{document}